\documentclass{article}

\usepackage[accepted]{icml2026}
\usepackage{microtype}
\usepackage{graphicx}
\usepackage{booktabs}
\usepackage{amsmath,amssymb}
\usepackage{enumitem}
\usepackage{hyperref}

\icmltitlerunning{Modeling Accessibility-Constrained Networks with Time-Weighted Graphs}

\begin{document}

\twocolumn[
\icmltitle{Modeling Accessibility-Constrained Networks with Time-Weighted Graphs}

\begin{icmlauthorlist}
\icmlauthor{Marc Walden}{ucla}
\icmlauthor{Jason Liu}{ucla}
\icmlauthor{Ryan Liu}{ucla}
\icmlauthor{Hamza Khan}{ucla}
\end{icmlauthorlist}

\begin{center}
{\small
$^{1}$University of California Los Angeles, Los Angeles, CA, USA
}
\end{center}

\printAffiliationsAndNotice{}

\vskip 0.3in
]

\begin{abstract}
Accessibility for the physically disabled is a prevalent issue on university campuses, where stairs and steep slopes make navigating campus arduous. Our work proposes a pipeline to model a college campus as a network by combining Strava and other APIs with depth-first search to derive insights into wheelchair-accessible paths. We then develop a custom Least Resistance algorithm to compute optimal paths between selected nodes and benchmark it against Dijkstra’s algorithm. We highlight crucial nodes on campus using centrality measures, demonstrating that wheelchair users are significantly constrained by the lack of mobility options and accommodations. Our pipeline is designed to support future expansion in scope and accuracy, while enabling the proposal of engineering solutions to improve campus accessibility for physically disabled individuals.

\end{abstract}

\section{Introduction}
Westwood is notoriously hilly and the University of California Los Angeles was subsequently designed with many stairs and steep pathways to maximize the terrain. This often makes the campus inaccessible to physically disabled students who attempt to navigate primarily through elevators or moderate inclines. Some integral halls and centers are entirely out of reach for wheelchair users and many more require lengthy detours. Physically disabled students comprise 2.1\% of the 48,000 students at UCLA. Moreover, the influx of students using wheeled transport, including electric scooters and skateboards, has made campus accessibility an increasingly salient issue. UCLA is one of the largest public universities, hosting countless clubs, sports events, international speakers, performing arts, and soon the 2028 Olympics and Paralympics. Accommodating these populations is necessary for UCLA to encourage attendance and visitation regardless of physical ability. Strides in this direction will also encourage campuses nationwide to reassess their accessibility. This project was also inspired by Salkhi Khasragh, who explored how pedestrian dynamics on university campuses can be modeled using complex network theory in her paper, “University Campus as a Complex Pedestrian Dynamic Network: A Case Study of Walkability Patterns at Texas Tech University (TTU)”~\cite{khasraghi2024campus}. Salkhi’s application of network theory and computation provides novel insights into the structure of TTU. The utility of networks and centrality measures enables us to derive insights into the accessibility of UCLA and propose necessary changes to improve campus accommodations.

\section{Background}
This project aims to identify areas of inaccessibility on UCLA's campus, and calculate optimal routes for wheeled access. We employ APIs and mapping software to model the main campus and residential halls and divide it into a complete network of nodes and edges. Edges represent pedestrian pathways while nodes represent the intersection of these paths. We note which paths are wheelchair-accessible based on their use of stairs and steep inclines. Steep inclines are legally regulated as slopes greater than an 8\% or 1:12 ratio of rise to run. Nodes are registered as geodesic latitudinal and longitudinal coordinates. These coordinates are measured in degrees which we convert to meters to measure distance. We assume that the curvature of the Earth is negligible. We calculate traversal times based on distance, speed, and elevation and use them as weights for the network edges. We design shortest path and least resistance algorithms to calculate the fastest and most efficient routes for disabled individuals for any given destination. The shortest path algorithm is an implementation of the traditional Dijkstra's algorithm, logging the travel time and nodes traversed. This enables us to depict where UCLA's campus struggles with wheel accessibility by analyzing large disparities in travel time. With this data, we can propose efficient alternatives that wheelchair users could leverage in lieu of stairs or steep inclines. The results of the project will promote inclusivity and equal opportunity for education regardless of physical ability.

\section{Data Collection}
\paragraph{Prototype Model.}
Our prototype model involves a manual process of data collection in which we picked nodes on campus based on arbitrary importance. We construct a map comprising 21 nodes that represent points of interest that branch into 3 or more paths. By connecting the nodes, we achieve 43 potential paths. We determine which paths are not wheelchair-accessible by determining if they require stairs or steep slopes, defined by inclines of 30 degrees or greater. This leaves 32 paths for comparison between a wheelchair user and a non-disabled individual. Figure 1 displays the final map of UCLA's campus with nodes representing key intersections and edges representing paths between them. Blue edges represent paths that are non-accessible for wheelchair users. 

\begin{figure}[htbp]
\centering
\includegraphics[width=0.9\linewidth, height=0.25\textheight]{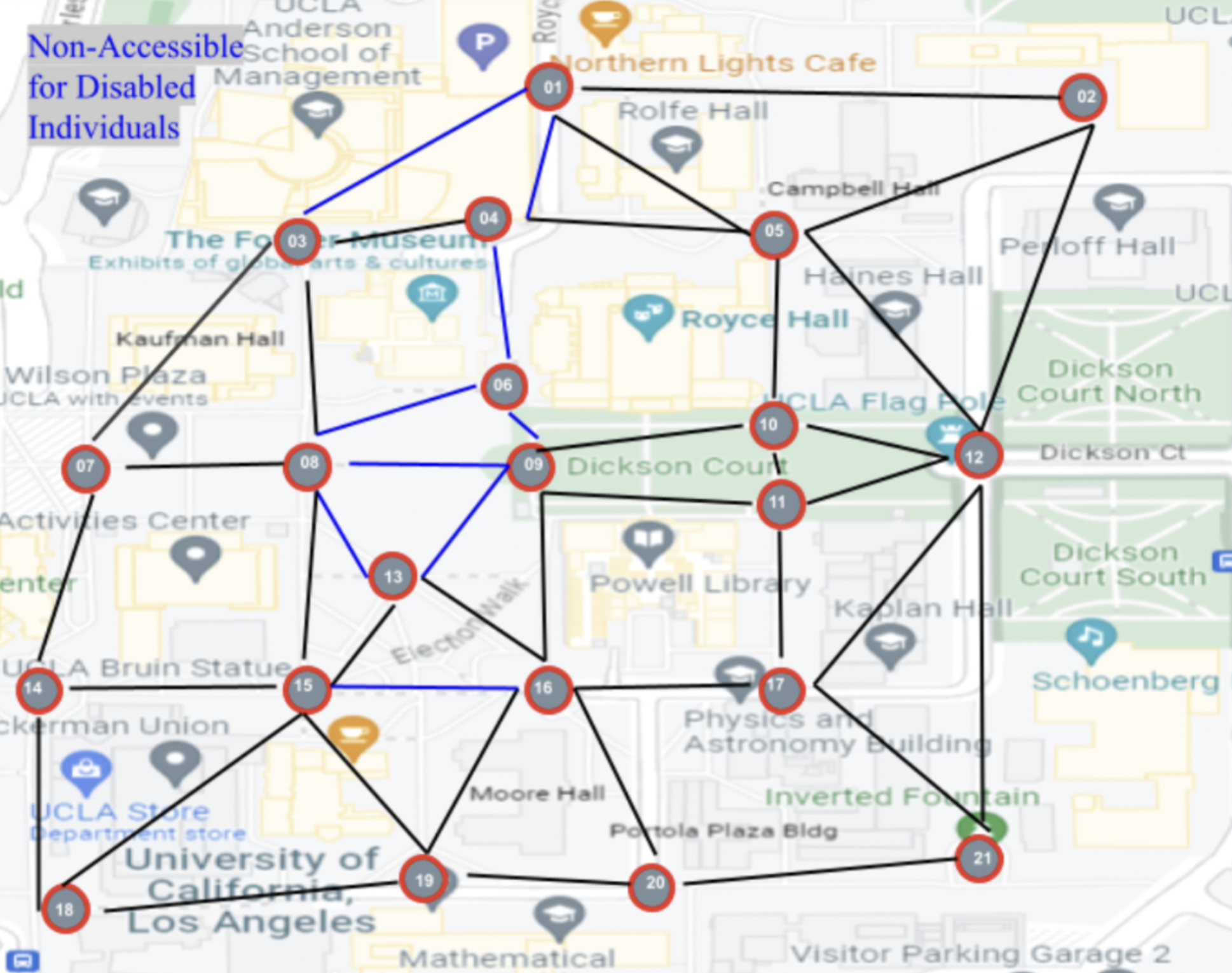}
\caption{Nodes and Edges for the Prototype Model}
\label{fig:1}
\end{figure}

Each path is then physically traversed by one of our four group members at least once by walking and once by simulating the use of a wheelchair ~\cite{maverick2023wheelchair}. Subsequently, the times are recorded and stored into matrices as shown in Figure 2. Each entry i, j represents the time taken to traverse the edge by one of our members. Both matrices for non-disabled and disabled traversal times are included.

 Initially, the main goal of our project was to use solely this method. However, we realized that this approach was prone to significant error and limitations due to several factors, including--but not limited to--human error, inconsistencies in walking speed among group members, physical fatigue, and scalability constraints as the project expanded. To address these challenges and improve the reliability of our data, we decided to elevate our data collection process using more systematic methods. 
 
 \paragraph{Production Model.}
 For our alternative method, we decided to employ APIs and mapping software to model the main campus and residential halls and divide it into a complete network of nodes and edges. We chose to use the Strava Global Heatmap because it was publicly callable via an API, and since it is based on collating user GPS information, it will include even minor or unofficial paths, along with what modes of transport they are navigable with. We were able to get an API key by making a free account through the website and imported it into JavaScript OpenStreetMap. From there, information was distilled into lines and imported into QGIS as vector line geometry in the WGS84 projection system. From there, we cropped the information to only include the reasonable scope needed to navigate UCLA's campus. We then ran commands to create nodes on all intersections of lines. From there, we specified the construction of all remaining vertices within the geometry, which essentially are points along each path that track the behavior of the path. So, if the path has curves, we will be able to track the curve instead of treating it as a straight line. After splitting the geometry along all these vertices, we essentially had a massive network of nodes, each connected by vertices and line segments, as we can see in Figure 3.

 \begin{figure}[H]
\centering
\includegraphics[width=0.9\linewidth]{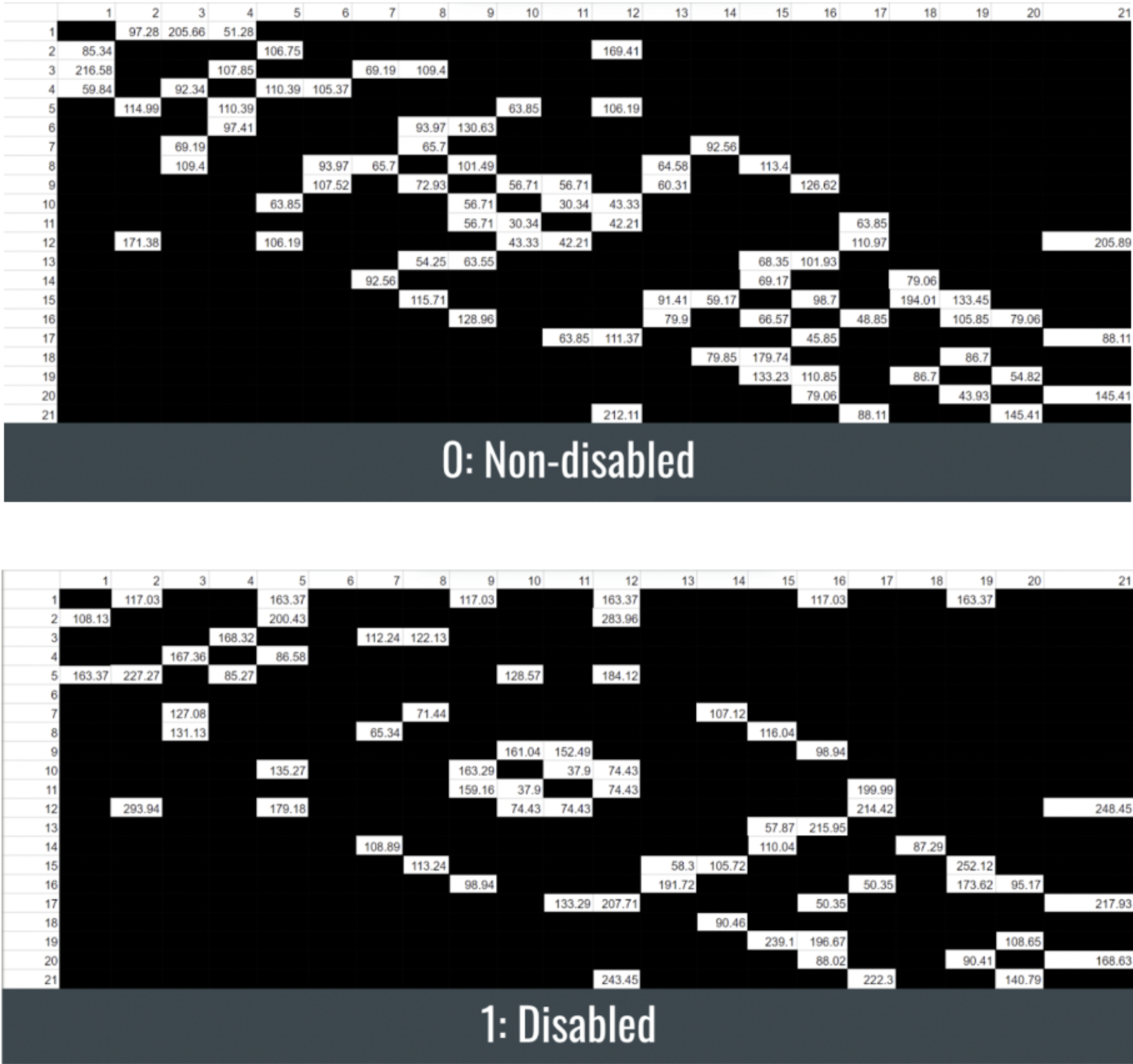}
\caption{Time taken for physical traversal between all pairs of nodes with an edge between them for disabled and non-disabled individuals}
\label{fig:2}
\end{figure}

 Then, we wanted to find a way to keep track of all the vertices traversed on a path between two nodes, so we developed a depth-first search algorithm to aid us in the process. Please note that, in the rest of the section, we will be referencing nodes as yellow nodes, and vertices as red nodes, matching the colors in Figure 3.

\begin{figure}[H]
\centering
\includegraphics[width=0.9\linewidth]{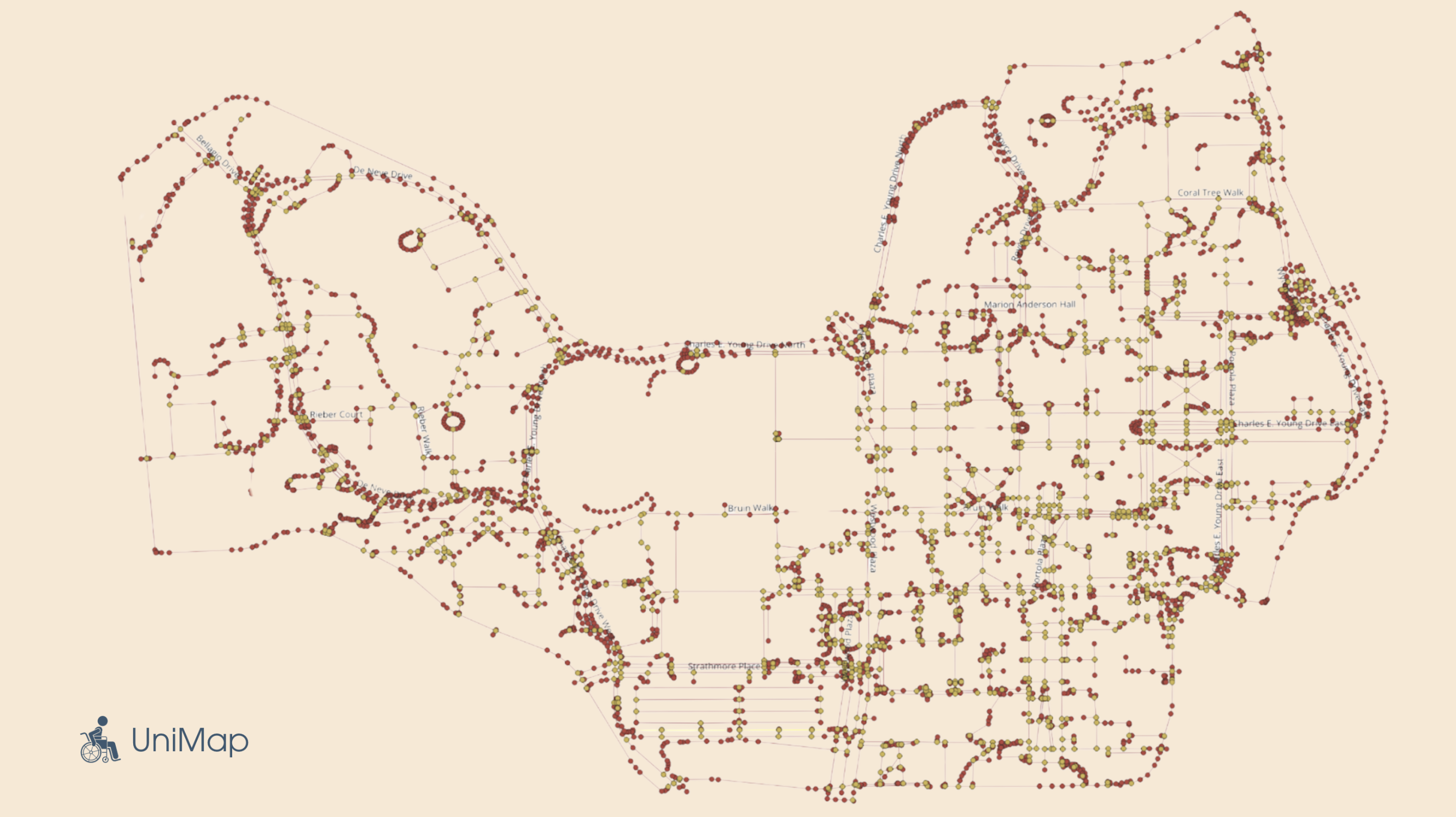}
\caption{Nodes, vertices, and edges after using Strava's API and Depth-First Search}
\label{fig:3}
\end{figure}

 Before being able to calculate the times it takes to traverse between two yellow nodes, we first needed to perform an intermediate step. The information stored for a given pair of yellow nodes is not just limited to the time it takes to traverse them. Instead, we needed to add intermediate nodes along each path to account for information like curvature, elevation changes, and deviations from straight-line travel that would otherwise be missed in a simplified graph. Each of the intermediary nodes stores their specific coordinates and elevation. By capturing these finer details, the model more realistically reflects the true effort or time required to traverse each path. For example, consider two yellow nodes with equal elevation connected by a curved path that passes over a hill. Although the start and end points are level, the actual route requires ascending and descending terrain. This particular situation poses a challenge: without incorporating intermediary nodes, the Dijkstra's and Least Resistance algorithms would simply detect an edge with no elevation change that can be traversed in a strictly straight line. As a result, the true physical effort or time cost would be underestimated, and the algorithm would also lack the necessary information to determine whether the path is truly accessible for disabled individuals. However, the integration of intermediary nodes that store coordinates and elevation changes breaks down complex paths into smaller segments and allows the algorithm to capture information that would otherwise be overlooked. These resulting path calculations more accurately reflect real-world travel conditions. Ultimately, their purpose is to bridge the gap between abstract graph representations and the physical realities of the environment. To control the spacing between these intermediate nodes, we introduced a tunable parameter $c$. This parameter determines the maximum allowed distance between two consecutive points where data (such as distance, slope, or elevation) is recorded. In order to maximize precision and obtain more detailed path representations, we generally aimed for relatively smaller values of c. However, smaller values of c also increase the number of nodes and edges in the graph, which can lead to higher computational costs during path finding. Therefore, we selected $c \approx 4.5 \times 10^{-5}$ in geodesic degrees (approximately 5 meters),
 providing a balanced trade-off between accuracy and efficiency.

\begin{figure}[H]
\centering
\includegraphics[width=0.9\linewidth]{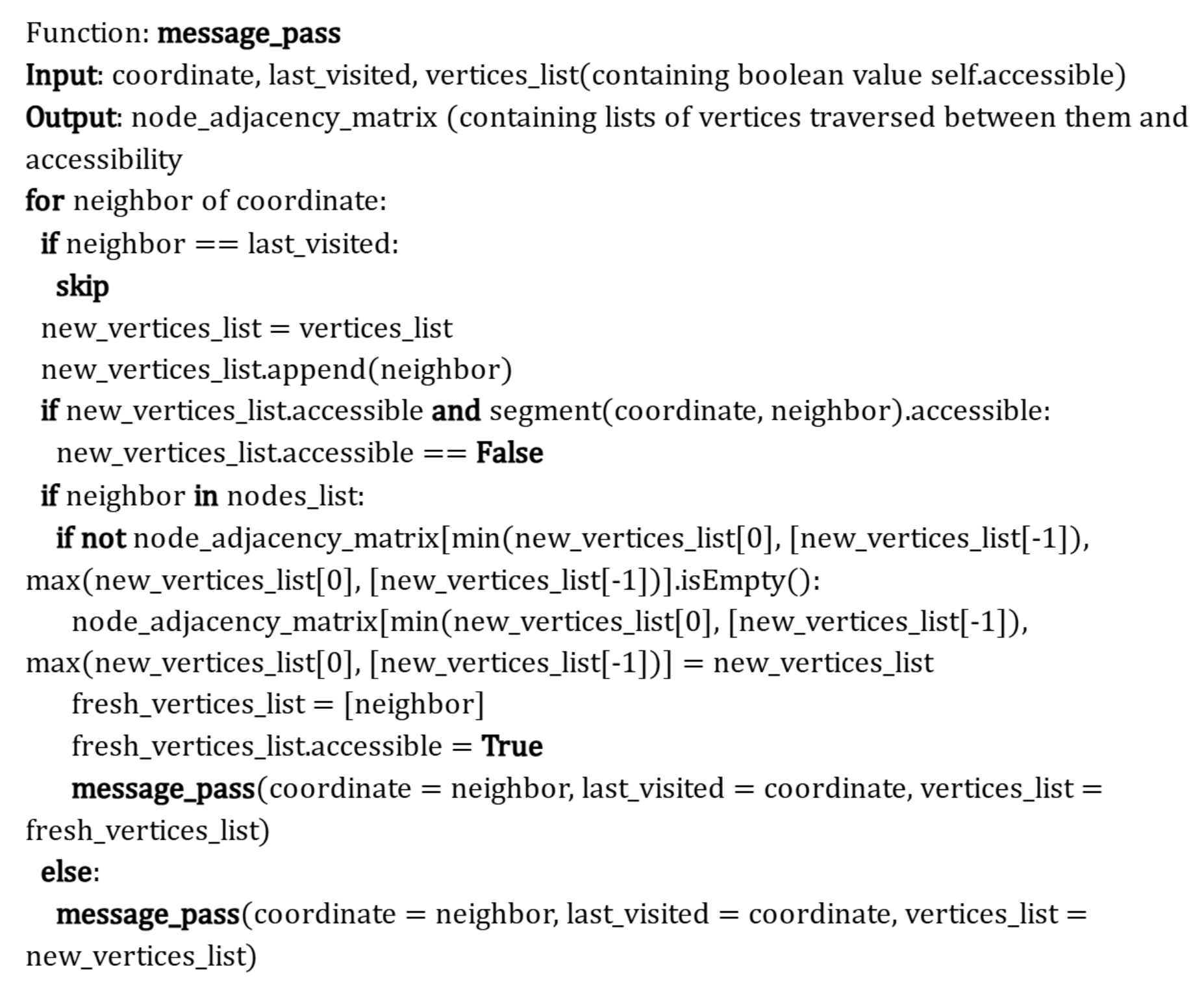}
\caption{Pseudocode for the Depth-First Search Algorithm}
\label{fig:4}
\end{figure}

\section{Methods and Models}

\paragraph{Dijkstra's Algorithm.}
Similarly to our two methods for data collection (prototype and production model), we also explore two different alternatives for finding the shortest time to travel between two given nodes and the corresponding path that needs to be taken. In this section, we firstly introduce Dijkstra's algorithm~\cite{kapoor2023dijkstra}, a popular method for computing the shortest path in weighted graphs with non-negative edge weights. Additionally, we introduce an original, custom-made Least Resistance algorithm, which simulates the flow of energy, or voltage, through a network to find the path of least resistance between two given nodes. The main idea behind Dijkstra's algorithm is to incrementally explore nodes and update the minimum travel time until the destination is reached. As discussed previously, we use travel time to measure edges, the connections between all N nodes, and map the shortest time to each node. The goal is to fill in an N $\times$ N matrix where entry i, j represents the time it takes to traverse from node i to node j. Note that this matrix is not perfectly symmetric because there can exist changes in elevation in the positive or negative direction between any two given nodes. Dijkstra's algorithm begins by initializing a matrix with 0’s along the diagonal representing zero distance from a node to itself and infinity everywhere else to indicate initially unknown distances. Starting from a designated source node, the algorithm stores a variable for the current known shortest time to a given node. This variable is updated whenever a shorter route is found through a neighboring node using a min-heap to efficiently select the next node with the smallest known time. Once the shortest path to a node is confirmed, that node is marked as visited and no longer considered for further updates. This process continues until all nodes have been visited, at which point the algorithm can return the shortest path and travel time between any user-specified destination and arrival node. After incorporating the min-heap structure, Dijkstra's algorithm’s time complexity is $O(E \log N)$, where $N$ is the total number of nodes and E is the total number of edges. This makes it a reliable and efficient choice for computing shortest paths in large, weighted graphs with non-negative edge weights.

\paragraph{Least Resistance Algorithm.}
The second method is a least resistance algorithm, which is inspired by the way that electricity travels through space. The algorithm simulates the flow of energy, which we term as voltage, and it gradually increases as the algorithm explores different paths. We include a boolean value to reflect the physical ability of the user, which indicates whether to use the data for disabled or non-disabled. The algorithm is centered around two sets of arrays that contain all adjacent nodes to the start and end nodes, respectively, showing all paths that can be completed within the current voltage (besides the initial values). From both the starting and ending nodes, the algorithm continuously explores all potential paths, considering only those where the energy required to travel is within the current voltage. Every time a path is traversed, the energy required to travel said path is subtracted from the voltage value, determining how much energy is left to use. As the algorithm runs, the voltage is incrementally increased, allowing a wider range of paths to become accessible over time. With each voltage level, the algorithm expands its collection of possible paths from the start and end nodes to various points on the graph. Eventually, assuming there exists a connection between the two nodes, these paths converge--either at an intermediate node or between two adjacent nodes. In the latter case, the algorithm checks whether the remaining voltage is sufficient to bridge the gap. Once a connection is found, the final route is constructed by combining the start and reversed end paths. Finally, the time it takes to traverse the route is calculated by finding the sum of the times between all the nodes in the route. Our Least Resistance algorithm differs from Dijkstra's algorithm in both structure and computational focus. While Dijkstra's algorithm relies on initializing and updating a full matrix or distance map to track the shortest distances from a source node to all other nodes in the graph, our approach avoids this overhead. Instead, the Least Resistance algorithm is designed to solve for a single, user-specified pair of nodes at a time, making it much more computationally cheap. It incrementally explores only the necessary portions of the graph, expanding outward from both endpoints until a viable connecting path is found. By forgoing the need to compute distances to all nodes, the algorithm reduces unnecessary computations and memory usage, making it significantly more efficient in scenarios where only point-to-point travel is required--which arguably represents the majority of practical applications! This targeted exploration allows for faster performance in graphs at a much larger scale than just UCLA's campus. 

\paragraph{Elevation.}
Wheelchair accessibility is primarily compromised by stairs and steep slopes. Steep slopes can be optimally identified by changes in elevation between intermediate nodes. Slopes must be at most 8\% incline to be legally wheelchair compliant. Elevation would also help tune the travel time by adjusting for downhill and uphill variables. We employed elevation APIs to get the appropriate data. The APIs take geodesic coordinate inputs and return the elevation in meters. The first API we attempted to use was Open Maps API~\cite{metterhausen2023elevation}. After calling the API for a portion of our network, we noticed discrepancies between the returned elevation and the actual elevation. The issue persisted so we pivoted to an alternative elevation API called Open Topo. Unfortunately, both tools failed to give accurate elevation data with the scale we were operating in. We were unable to find a feasible solution within the project duration and intend to continue pursuing different options in the future. Computing elevation changes on the 5 meter intervals will enable us to refine the set of wheelchair-accessible edges and reduce the error for travel time. 

\paragraph{Centralities.}
Centrality measures are one of the primary utilities of networks. For the case of our transportation network, we considered three centralities including closeness, betweenness, and Katz. Closeness centrality is computed by the inverse of the average shortest distance between the vertex and all other vertices in the network. It favors nodes that are closest to the most number of other nodes in distance. While this metric may be effective to identify highly connected locations on campus, it doesn’t offer much benefit for our purposes. Katz centrality takes into account the amount of direct connections a node has as well as the relative importance of the nodes it’s connected to. This measure is particularly holistic but again offers little utility in the case of identifying inaccessible areas of campus or other locations of issue on campus. Betweenness centrality is the main measure we employ in our network. It is computed by calculating the number of shortest paths a node appears on over the total number of shortest paths. This means nodes with high betweenness are commonly used on the shortest paths. In the context of our network, this helps us identify which nodes are crucial for wheelchair-accessible pathways, and which are likely bottlenecks or risk of high traffic. Nodes with high betweenness also help draw attention towards parts of the network that would most benefit from new construction for a path that can reduce average travel time. We computed betweenness centrality using the NetworkX package. We have our weighted matrix with the travel times between each intersection and generate the resulting betweenness measures for each node. As we continue expanding the network and labeling the true wheelchair-accessible edges using the elevation API, betweenness centrality will offer novel insight into nodes of importance on campus.

\section{Results}
Since our prototype model only took into account a subsection of UCLA's campus, we only accounted for 21 different nodes, which gave us a 21 $\times$ 21 matrix of times it takes to traverse a path between two directly connected nodes, which we can see in Figure 1. However, for this project, we expanded our scope to the entirety of UCLA, so our output matrix had dimensions 1290 $\times$ 1290. We ran our Dijkstra's algorithm and our least resistance algorithm as a method to double-check our results. Dijkstra’s algorithm runs for all possible nodes in the network and only ends when all nodes (assuming fully connectedness) have a value for the time. But after running the algorithm one time, the matrix of times between any two nodes in the network is now obtained, so we then decided to use our least resistance algorithm to confirm our values. The least resistance algorithm is more efficient than Dijkstra's because it can find the shortest path between two nodes without computing values for all the other nodes, contrary to Dijkstra's algorithm. So, to ensure our times from Dijkstra's algorithm were accurate, we double-checked 100 random node pairs using our least resistance algorithm and were able to attain matching results, hence proving the accuracy of our model. Although we cannot provide a visualization to showcase this verification process, successfully scaling our model from 21 to 1,290 nodes and confirming its accuracy through two independent algorithms stands as one of the most significant accomplishments of our project. Therefore, we were able to successfully analyze the travel times for disabled and non-disabled people to evaluate places of need for infrastructure development. Firstly, as we see in Figure 5, the average time of a disabled person is much higher, almost double that of a non-disabled person. This can be attributed to a number of factors, such as certain paths that are almost entirely inaccessible or extremely long, windy detours that disabled people must take to reach a certain location. One prime example is the path it takes to go from Kerckho Hall to the Mathematical Sciences building, which requires wheelchair users to traverse via winding ramps (see Figure 1). This increases the time taken by a factor of 2.7 compared to non-disabled people, who only need to climb a set of stairs to reach the building. Another example is the path from Fowler Museum to Northern Lights Cafe, where large detours and u-turns are needed. Wheelchair users aiming to reach Northern Lights Cafe from the area around Rolfe Hall face significant challenges due to the lack of ramps or slopes ~\cite{ufmi2023mapucla}. To reach higher elevation, they must use an elevator inside Rolfe Hall, as there are no accessible paths that bypass the building. Alternatively, users are forced to take a lengthy detour to Bunche Hall, make a U-turn, and backtrack to the cafe--significantly increasing travel time and effort. Simple shallow ramps or signs to help inform students or visitors of nearby elevators could be added on campus to help accelerate these travel times. \begin{figure}[H]
\centering
\includegraphics[width=0.9\linewidth]{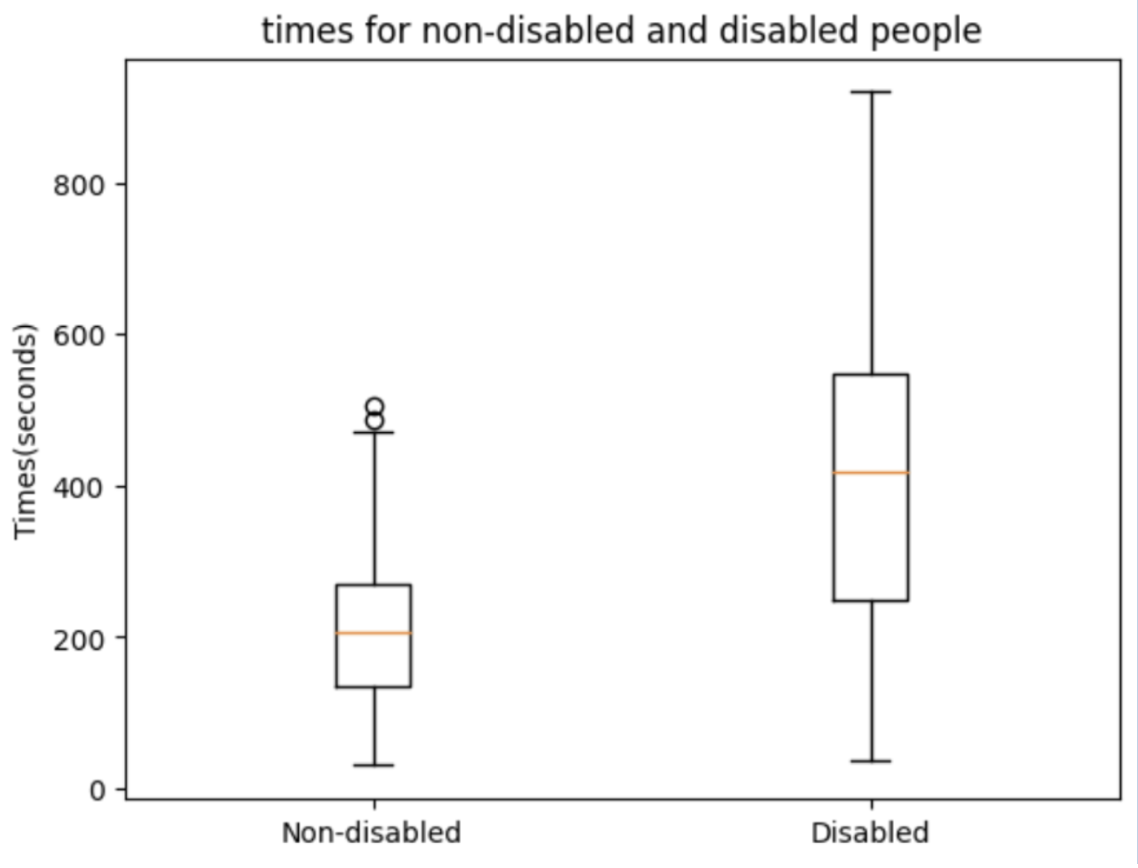}
\caption{Boxplot for times taken to traverse between node pairs.}
\label{fig:5}
\end{figure}
 We were also able to determine betweenness centralities for our network to highlight nodes that are commonly used in shortest paths. This analysis provides valuable insights into which parts of campus experience high traffic and which nodes are crucial for maintaining wheelchair accessibility. Betweenness centrality measures the frequency at which a node appears on the shortest path between pairs of nodes. Nodes with higher betweenness centrality play a pivotal role in maintaining connectivity across the network, as they act as essential intermediaries between various paths. Our results revealed that the locations on campus with the three highest betweenness centralities were Wilson Plaza, Pauley Walk, and the John Wooden Center. These findings align with our expectations, as these locations are well-known hubs of movement and social interaction on campus. Pauley Walk and Wooden Center consistently show high centrality values due to their strategic positions between the residential areas on the Hill and the academic core of campus. Whether it’s students going to and from the Hill, heading to the gym, or going to class, these two locations serve as crucial intersections on campus. Wilson Plaza connects major academic buildings and is frequently used as a passage between the northern and southern parts of campus. Its central location makes it a vital link for anyone traversing from one side to the other, and as a result, it exhibits high betweenness centrality. As we see in Figure 5, a small number of often traversed nodes reaches a maximum betweenness centrality of around 0.175. We see higher frequencies of nodes with smaller betweenness centralities, indicating that the majority of campus locations are less frequently utilized as connection points between paths. These nodes typically correspond to localized areas or peripheral locations that are not essential for traversing the campus as a whole. If any of these high betweenness centrality nodes were obstructed or closed for maintenance, it could significantly disrupt mobility and force individuals, particularly those with disabilities, to take significantly longer and less convenient routes. By identifying these high-centrality nodes, campus planners can prioritize accessibility improvements, ensure regular maintenance, and develop contingency plans to minimize disruptions in campus mobility. \begin{figure}[H]
\centering
\includegraphics[width=0.9\linewidth]{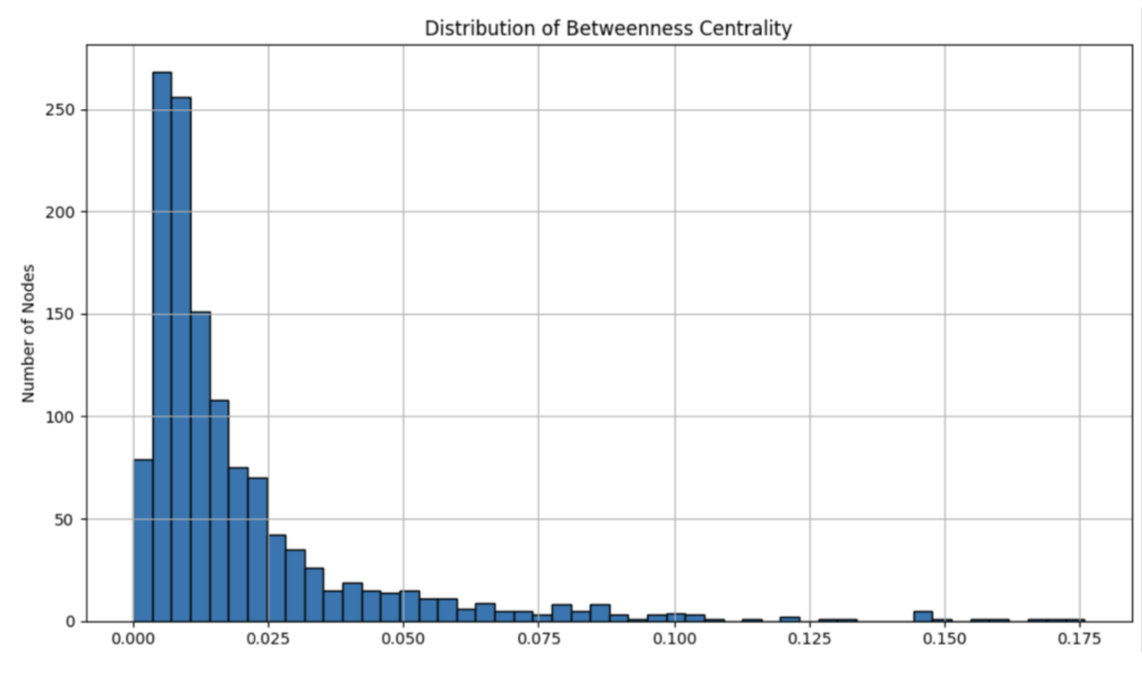}
\caption{Distribution of Betweenness Centralities}
\label{fig:6}
\end{figure}

\section{Conclusion}
UCLA struggles with campus accessibility for the physically disabled. We modeled the university with a network, dividing it into nodes and edges that represent pedestrian pathways and intersections. We weight the network with the travel times for each node and use shortest path algorithms to identify optimized pathways. We employ centrality measures to highlight key locations and potential engineering solutions. We struggled with the integration of elevation APIs which limited the insights we could draw. However, we were very successful in producing a comprehensive network of campuses with accurate travel times that enable us to find optimal paths for physically disabled individuals. For our future steps, we aim to enhance our model by incorporating elevation data through a reliable and accurate elevation API. Elevation plays a crucial role in determining the accessibility of pathways, especially for individuals who use wheelchairs or have limited mobility. Integrating elevation data will not only increase the precision of our model but also enable us to evaluate the incline of each path to ensure it remains within the maximum allowable incline of 8\%, as specified by ADA (Americans with Disabilities Act) standards. This addition will significantly improve our ability to identify paths that are not wheelchair-friendly and guide users toward more accessible routes, as well as more accurately calculate the travel times for all the paths. We would like to create some sort of software that can detect locations that are primed for potential infrastructure changes, since we currently do not have an algorithm to propose new design changes on campus. Developing an algorithm that can analyze traffic patterns, betweenness centralities, and accessibility metrics could help identify bottlenecks, high-traffic zones, or steep inclines that would benefit from improvements. By pinpointing locations where new paths, ramps, or better signage could enhance accessibility, we can assist campus planners in making data-driven decisions. Finally, we hope to expand the scope of this project into other universities, since UCLA is not the only campus with accessibility issues. Our long-term goal is to create a comprehensive tool that empowers institutions to make their campuses more inclusive and accessible to all students and visitors.

\bibliographystyle{icml2026}
\bibliography{references}

\end{document}